\documentclass[pre,twocolumn,showpacs]{revtex4}
\usepackage{graphicx,amsmath,amssymb}
\newcommand{\figwidth}{3.2in}

\begin{document}

\title{Local Elastic Constants in Thin Films of an FCC Crystal}

\author{Kevin \surname{Van Workum}}
\affiliation{Department of Chemical Engineering, University of
Wisconsin-Madison}

\author{Juan J. \surname{de Pablo}}
\affiliation{Department of Chemical Engineering, University of
Wisconsin-Madison}

%\date{\today}

\begin{abstract}
In this work we present a formalism for the calculation of the
local elastic constants in inhomogeneous systems based on a method
of planes. Unlike previous work, this formalism does not require
the partitioning of the system into a set of finite volumes over
which average elastic constants are calculated. Results for the
calculation of the local elastic constants of a nearest neighbor
Lennard-Jones fcc crystal in the bulk and in a thin film are
presented. The local constants are calculated at exact planes of
the (001) face of the crystal. The average elastic constants of
the bulk system are also computed and are consistent with the
local constants. Additionally we present the local stress profiles
in the thin film when a small uniaxial strain is applied. The
resulting stress profile compares favorably with the stress
profile predicted via the local elastic constants. The surface
melting of a model for argon for which experimental and simulation
data are available is also studied within the framework of this
formalism.
\end{abstract}

\pacs{68.08.-p,68.35.Gy,68.60.-p}

\keywords{elastic constants, thin films, local elasticity}

\maketitle

\section{Introduction}

Knowledge of the local elastic constants in inhomogeneous systems
is of significant theoretical, experimental, and industrial
interest. As nanofabrication technologies improve and allow for
the design and construction of nanoscopic devices, understanding
the mechanical response of materials at nanometer length scales
will become increasingly important. In particular, deviations from
bulk, continuum behavior may lead to complications in the
manufacturing of such devices. For example, in the
microelectronics industry, the mechanical collapse of photoresist
structures below 100nm may limit the ultimate density of memory
storage devices or the performance of
microprocessors~\cite{tanaka93,cao00,boehme02}.

In nanoscopic structures, interfaces are likely to play a major
role in apparent deviations from bulk, continuum
behavior~\cite{fryer2}. The interface could either weaken or
reinforce the overall mechanical behavior of the structure,
depending on the nature of interactions between adjacent domains
and the size of the structure. Understanding how mechanical
properties vary near interfaces or free surfaces would provide
insights into such phenomena.

Knowledge of interfacial behavior is crucial for understanding the
adhesion of thin polymer films, where the inter-diffusion of the
polymers and the molecular mobility near the film boundaries play
a significant role~\cite{torres00}. Properties such as adhesion,
dewetting, and surface melting in thin films are likely to be
controlled by processes that occur within the first few nanometers
of the interface. It would therefore be beneficial to have the
ability to measure (computationally or experimentally) physical
properties with molecular spatial resolution.

A microscopic definition for local elastic constants has been
proposed in the literature~\cite{lutsko88,eerden90}.
Implementation of that formalism requires that
\textit{layer-averaged} local elastic constants be determined. For
inhomogeneous systems, the results from averaging over a
particular layer depend strongly on the size and position of that
layer. This is particularly true in an interfacial region or near
a free surface, where material properties can change significantly
over short distances.

In this work, we are interested in the local elastic constants and
surface melting of thin crystalline films. Specifically, we
present a formalism in which the local elastic constants are
calculated at precise planes in the system, as opposed to small
volumes or slabs. In the bulk, the calculated local elastic
constants are verified by averaging over the entire system and
comparing the results to the bulk value. The local elastic
constants in the film are verified by comparing the local stress
profiles that arise from uniaxial strain and those calculated
directly from the elastic constants.

\section{Theory}

In a homogeneous material, applying a homogeneous strain
necessarily results in a homogeneous stress. The stress is given
by
\begin{equation}\label{thin:eqn:bulk_stress}
\sigma_{ij} = C_{ijkl} \epsilon_{lk},
\end{equation}
where $C_{ijkl}$ is the bulk elasticity tensor, and where the
indices represent the cartesian coordinates in three dimensions.

When a homogeneous strain is applied to an inhomogeneous system,
the resulting stress is also inhomogeneous. The local stress is
then given by
\begin{equation}\label{thin:eqn:local_stress}
\sigma_{ij}(\mathbf{r}) = C_{ijkl}(\mathbf{r}) \epsilon_{lk},
\end{equation}
where $C_{ijkl}(\mathbf{r})$ is the \textit{local} elasticity
tensor. The relationship between the local and bulk elasticity
tensors can be written as
\begin{equation}\label{thin:eqn:bulk-local}
C_{ijkl} = \frac{1}{V} \int_V C_{ijkl}(\mathbf{r}) d\mathbf{r},
\end{equation}
where $V$ is the volume of the system.

The bulk elasticity tensor can be expressed in terms of the
fluctuations of stress according to~\cite{ray82:_fluct}
\begin{equation}\label{thin:eqn:stress-stress}
 \begin{split}
   C_{ijkl} &= 2\rho k_BT \left[\delta_{il} \delta_{jk} +
   \delta_{ik} \delta_{jl} \right] \\
   &- \frac{V}{k_B T}\left[\langle P_{ij} P_{kl} \rangle - \langle
   P_{ij} \rangle \langle P_{kl} \rangle \right] + B_{ijkl},
 \end{split}
\end{equation}
where $\rho$ is the density, $\delta_{ij}$ is the Kronecker delta,
$P_{ij}$ is the pressure tensor and $B_{ijkl}$ is the so-called
Born term. The pressure tensor is given by
\begin{equation}\label{thin:eqn:pressure_tensor}
P_{ij} = \frac{1}{V} \left[ \sum_a p_{a_i}p_{a_j} / m_a -
\sum_{a<b} r^{-1}_{a b} u'_{ab} r_{ab_{i}} r_{ab_{j}} \right].
\end{equation}
The potential energy between interaction sites $a$ and $b$ is
denoted by $u_{ab}$, $r_{ab}$ is the distance between them, $p_a$
and $m_a$ are the momentum and mass of site $a$ respectively, and
the prime indicates a derivative with respect to $r_{ab}$. The
Born term is related to the first and second derivatives of the
potential energy of interaction by
\begin{equation}\label{thin:born}
B_{ijkl} = \frac{1}{V} \left \langle \sum_{a<b} \left[
\frac{u''_{ab}}{r^2_{ab}} - \frac{u'_{ab}}{r^3_{ab}} \right]
r_{ab_i} r_{ab_j} r_{ab_k} r_{ab_l} \right \rangle.
\end{equation}

In this work we focus on the elastic properties of thin films
having a planar symmetry; the films are assumed inhomogeneous only
in the direction perpendicular to the film, i.e. $z$.
Equation~(\ref{thin:eqn:stress-stress}) must therefore be modified
to calculate the elasticity tensor at precise planes within the
system, without need for bins or small volumes. To this end, we
use the method of planes (MOP)~\cite{todd95:_press} and obtain an
expression for the local elasticity tensor.

The first term in Eqn. (\ref{thin:eqn:stress-stress}) is the ideal
gas contribution to the elasticity tensor. The kinetic energy is
homogeneously distributed, even in inhomogeneous systems, and the
temperature is independent of $z$. However, the density can vary
in the $z$ direction. The density profile, $\rho(z)$, could be
calculated by dividing the system into many small bins and
counting the average number of particles per unit volume in the
bins. The density would then explicitly depend on the size of the
bins used. Alternatively, one can use the fact that for a free
standing film the total normal pressure, $P_{zz} = \rho(z)k_BT +
P^u_{zz}(z)$, is constant throughout the
system~\cite{varnik00:_molec}. In vacuum, we then have for the
density profile
\begin{equation}\label{thin:eqn:rho_z}
\rho(z) = -\frac{P^u_{zz}(z)}{k_BT},
\end{equation}
where $P^u_{zz}(z)$ is the configurational contribution to the
local pressure tensor. The local pressure tensor is the sum of
ideal and configurational terms, and can be calculated according
to~\cite{irving50, varnik00:_molec}
\begin{equation}\label{thin:eqn:MOP_pressure}
 \begin{split}
   P_{ij}(z) = &\rho(z) k_B T - \frac{1}{A} \biggl \langle \sum_{a<b}
   \frac{r_{ab_i} r_{ab_j}}{r_{ab}} u'(r_{ab}) \\
   &\times \frac{1}{|z_{ab}|}  \Theta \left( \frac{z-z_a}{z_{ab}} \right) \Theta
   \left(\frac{z_b-z}{z_{ab}}\right)\biggr \rangle,
 \end{split}
\end{equation}
where $A$ is the cross-sectional area of the film and $\Theta$ is
the Heaviside step function. The first term in Eqn.
(\ref{thin:eqn:stress-stress}) can then be written for
inhomogeneous systems as
\begin{equation}\label{thin:eqn:id_term}
C^{id}_{ijkl}(z) = 2 \rho(z) k_BT \left[\delta_{il} \delta_{jk} +
\delta_{ik} \delta_{jl} \right].
\end{equation}

The second term in Eqn. (\ref{thin:eqn:stress-stress}) arises from
bulk stress fluctuations; it accounts for the non-zero temperature
contribution to the elastic constants. We are interested in
relating the local stress, $\sigma(z)$, to a bulk homogeneous
strain. Therefore, instead of including the bulk-stress
bulk-stress correlation, we use the correlation between the
local-stress and the average bulk stress. The second term can then
be written as
\begin{equation}\label{thin:eqn:fluct_term}
C^{fluc}_{ijkl}(z) = - \frac{V}{k_BT}  \left[ \langle P_{ij}(z)
P_{kl} \rangle - \langle P_{ij}(z) \rangle \langle P_{kl} \rangle
\right].
\end{equation}
Note that the volume $V$ in Eqn. (\ref{thin:eqn:fluct_term})
cancels that in Eqn. (\ref{thin:eqn:pressure_tensor}) and there
also is no explicit volume dependence in the MOP expression for
$P_{ij}(z)$.

The Born term, Eqn. (\ref{thin:eqn:stress-stress}), can be
calculated at planes using the MOP in the same way the local
stress is determined, i.e. Eqn. (\ref{thin:eqn:MOP_pressure}). We
have for the Born term in inhomogeneous systems
\begin{equation}\label{thin:eqn:born_z}
 \begin{split}
B_{ijkl}(z) &= \frac{1}{A} \biggl \langle \sum_{a<b} \left[
\frac{u''_{ab}}{r^2_{ab}} - \frac{u'_{ab}}{r^3_{ab}} \right]
\frac{1}{|z_{ab}|} \Theta \left(
\frac{z-z_a}{z_{ab}} \right) \\
&\times \Theta \left(\frac{z_b-z}{z_{ab}}\right) r_{ab_i} r_{ab_j}
r_{ab_k} r_{ab_l} \biggr \rangle.
  \end{split}
\end{equation}
As before, this expression does not depend on the volume of the
system or the (arbitrary) size of a bin.

The final expression for the local elasticity tensor in
inhomogeneous systems with planar symmetry is given by
\begin{equation}\label{thin:eqn:local_C}
C_{ijkl}(z) = C^{id}_{ijkl}(z) + C^{fluc}_{ijkl}(z) + B_{ijkl}(z).
\end{equation}
We emphasize again that this expression for $C_{ijkl}(z)$ is valid
for inhomogeneous systems and is an average only over a cross
section (a plane) of the system, not a discrete volume. It
therefore relates the local stress (at $z$) to a homogeneous
strain. Also note that Lutsko et al.~\cite{lutsko88} have
presented a derivation for the local elasticity tensor, but they
averaged over a sub-volume in order to facilitate the
computations. It can be seen that by integrating over the entire
system, one recovers the bulk elasticity tensor, Eqn.
(\ref{thin:eqn:stress-stress}). We also note that this expression
does not require the use of any dynamic variables but only
requires ensemble averages taken from system configurations. It
therefore is useful in either molecular dynamics or Monte Carlo
(MC) simulations.

We note that other valid definitions of the local stress tensor
have been presented~\cite{tsai79,rowlinson82,harasima58} and
discussed extensively in the
literature~\cite{varnik00:_molec,hafskjold02}. These definitions
would in principle lead to different expressions for the local
elasticity tensor. Regardless of the definition, one should expect
to recover the bulk elasticity tensor after averaging over the
entire system. The definition used in this work is that of Irving
and  Kirkwood~\cite{irving50}. This definition was chosen here
because it has been shown to be a physically valid stress
tensor~\cite{hafskjold02} and it can be used in MC simulations.

\section{Simulations}

To demonstrate the calculation of the local elasticity tensor, we
employ the widely used nearest-neighbor Lennard-Jones (NNLJ) fcc
crystal model~\cite{parrinello81:_polym, fay92:_monte,
sprik84:_secon, cowley83:_some, li92:_fluct, ray85:_molec}. In
what follows, all results will be reported in dimensionless
Lennard-Jones units.

A bulk system consisting of 1000 particles with periodic boundary
conditions in all three dimensions was investigated first. This
system was simulated in the NVT ensemble at a temperature of $T=
0.3$ using a simple MC method. The density was chosen such that
the average bulk pressure is zero. The center of mass of one
atomic layer was held fixed at $z=0$. The average bulk elastic
constants for this system have been calculated previously and are
listed in Table~\ref{thin:data}. Reported elastic constants,
stresses and strains are represented using Voigt
notation~\cite{nye84:_physic}. For bulk fcc systems, there are
three groups of non-zero, independent elements of the elastic
constant matrix
\begin{equation}
C = \begin{bmatrix}
    C_{11} & C_{21} & C_{21} & 0 & 0 & 0 \\
    C_{21} & C_{11} & C_{21} & 0 & 0 & 0 \\
    C_{21} & C_{21} & C_{11} & 0 & 0 & 0 \\
    0 & 0 & 0 & C_{44} & 0 & 0 \\
    0 & 0 & 0 & 0 & C_{44} & 0 \\
    0 & 0 & 0 & 0 & 0 & C_{44}
  \end{bmatrix}.
\end{equation}

Second, we also consider a free standing film of 450 particles.
The free surfaces correspond to the $(001)$ face of the fcc
crystal. This system was also simulated in the canonical ensemble
using a conventional MC method. In this case the cross sectional
area was held constant with the same dimensions as the bulk
system. The film had nine atomic layers parallel to the free
surfaces. The temperature was the same as in the bulk, i.e. $T =
0.3$. The center of mass of the film was held fixed at $z=0$. For
an fcc film with free surfaces normal to the $z$-axis, there are
six groups of non-zero, independent elements of the elastic
constant matrix
\begin{equation}
C = \begin{bmatrix}
    C_{11} & C_{21} & C_{31} & 0 & 0 & 0 \\
    C_{21} & C_{11} & C_{31} & 0 & 0 & 0 \\
    C_{31} & C_{31} & C_{33} & 0 & 0 & 0 \\
    0 & 0 & 0 & C_{44} & 0 & 0 \\
    0 & 0 & 0 & 0 & C_{44} & 0 \\
    0 & 0 & 0 & 0 & 0 & C_{66}
  \end{bmatrix}.
\end{equation}

Local properties of the film and the bulk system were calculated
from Eqn. (\ref{thin:eqn:local_C}) at planes of constant $z$, with
each plane being separated by a distance of $0.02$ in both the
thin film and bulk systems. The average elastic constants were
also calculated in the bulk system from Eqn.
(\ref{thin:eqn:stress-stress}).

An additional simulation of the thin film was performed in which a
homogeneous, tensile, uniaxial strain was applied in the
$x$-direction, $\epsilon_1 = 0.01005$. The strain is defined
as~\cite{love27}
\begin{equation}
\epsilon_1 = \frac{1}{2} \left[ \left( \frac{L_x}{L^0_x} \right)^2
- 1 \right],
\end{equation}
where $L_x$ is the length of the simulation cell in the
$x$-direction, and $L^0_x$ is its original length. Since the
strain is homogeneous, it is known that the average strain in a
plane of atoms parallel to the free surface is equal to the
applied strain~\cite{kluge90}. The resulting stress profiles were
then calculated using Eqn. (\ref{thin:eqn:MOP_pressure}). The
stress profiles were also calculated directly from the elastic
constants using Eqn. (\ref{thin:eqn:local_stress}).

\section{Results}

\begin{figure}
  \centering
  \includegraphics*[width=\figwidth]{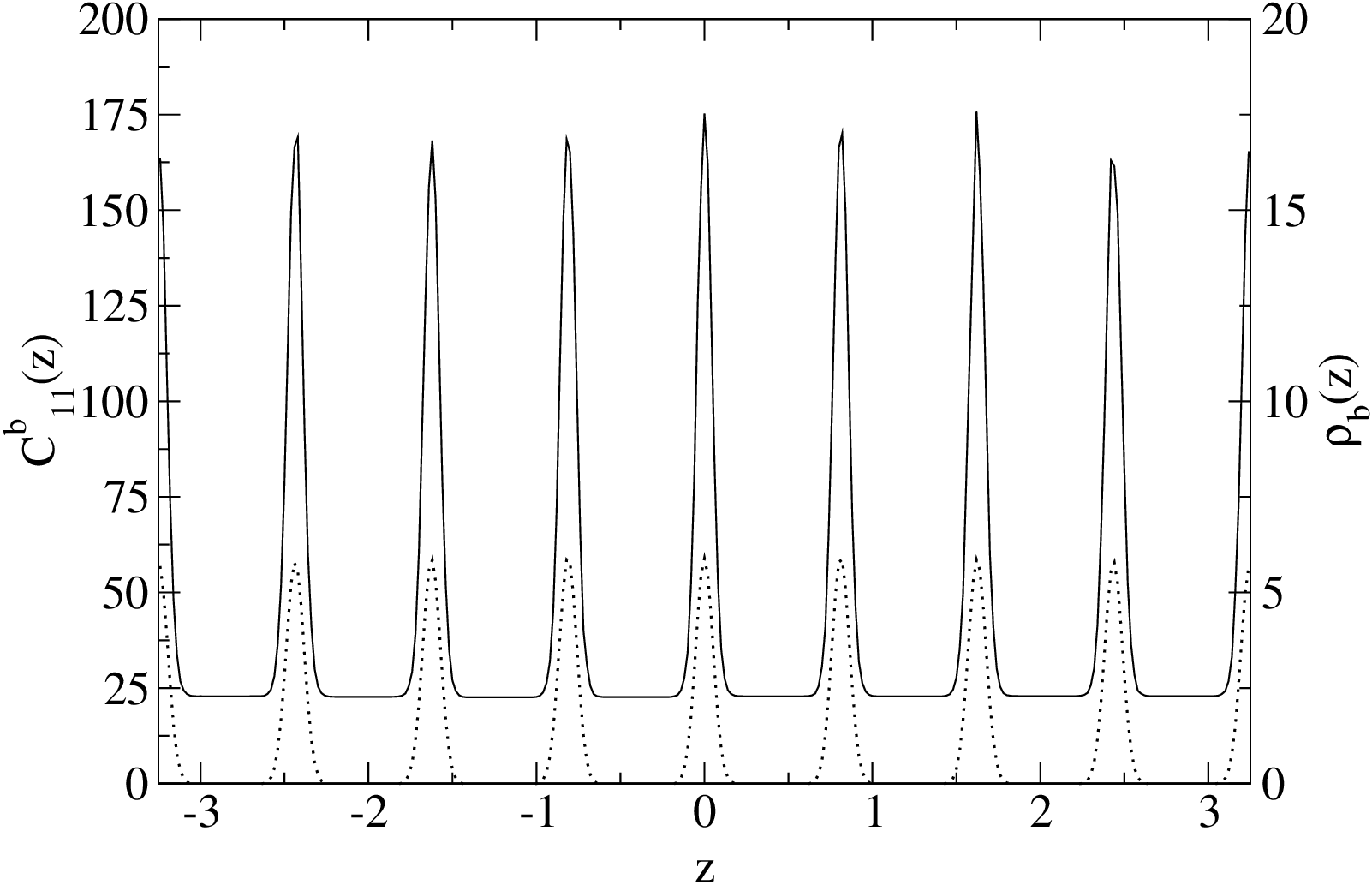}
  \caption{$C_{11}^b(z)$ (solid line) as a function of $z$ for
    the bulk system from Eqn. (\ref{thin:eqn:local_C}). The density profile,
    $\rho_b(z)$, is shown as the dotted line.}
  \label{thin:fig:C11bulk}
\end{figure}

\begin{figure}
  \centering
  \includegraphics*[width=\figwidth]{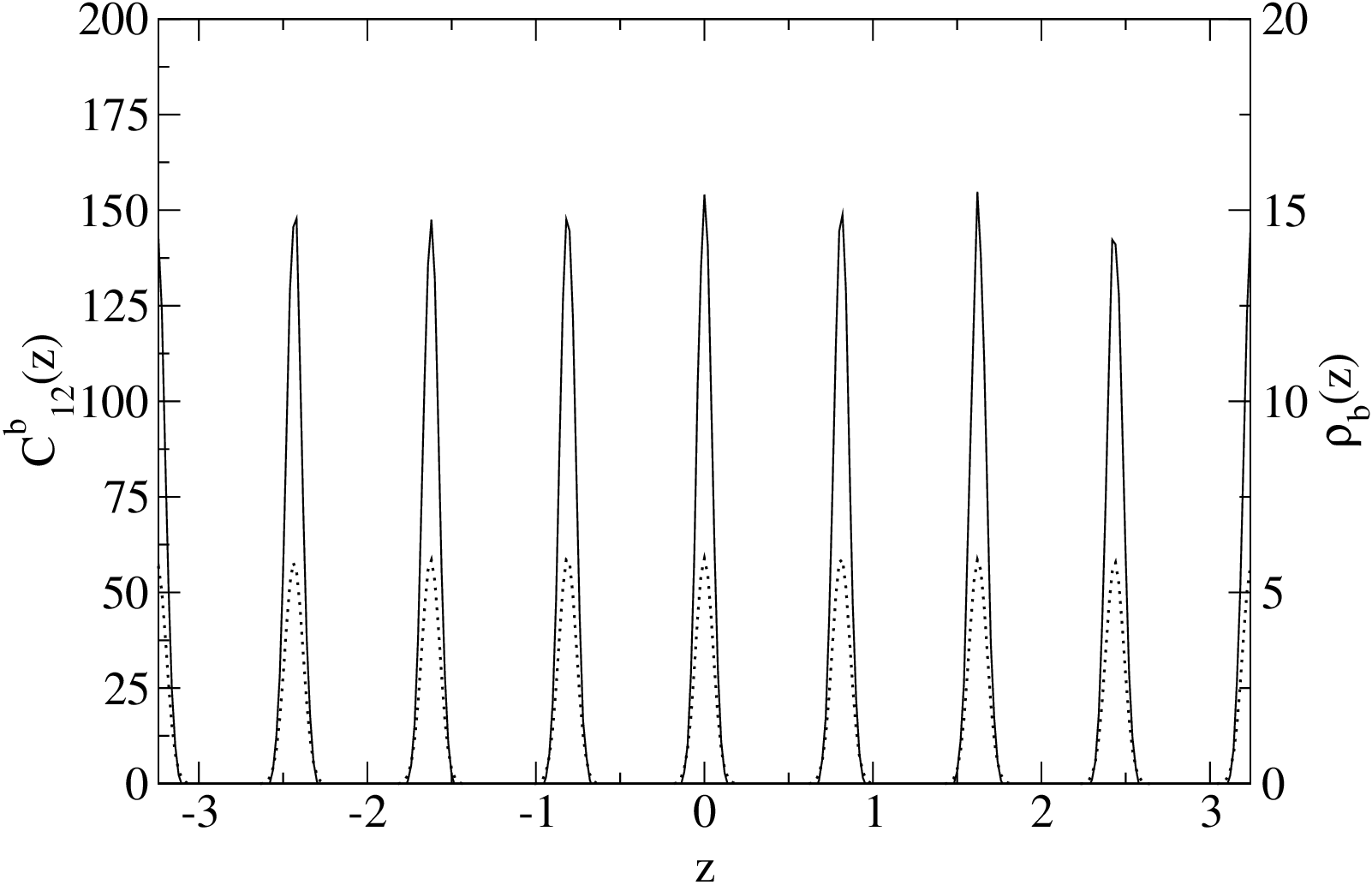}
  \caption{$C_{21}^b(z)$ (solid line) as a function of $z$ for
    the bulk system from Eqn. (\ref{thin:eqn:local_C}). The density profile,
    $\rho_b(z)$, is shown as the dotted line.}
  \label{thin:fig:C21bulk}
\end{figure}

The local elastic constant profiles for $C^b_{11}(z)$ and
$C^b_{21}(z)$ in the bulk system are shown in
Fig.~\ref{thin:fig:C11bulk} and Fig.~\ref{thin:fig:C21bulk}
respectively. The density profile, $\rho_b(z)$, is also shown in
these figures. Each peak in the profile at $C^b_{11}\approx175$
and $C^b_{21}\approx150$ corresponds to the center of mass of each
atomic layer. Each minimum at $C^b_{11}\approx22.5$ and
$C^b_{21}\approx0$ corresponds to the midpoint between each atomic
layer. The local elastic constant profile for $C^b_{44}$ is
similar to $C^b_{21}$ and is not shown.

\begin{table}
\centering
\begin{ruledtabular}
\begin{tabular}{cccc}
$C_{ij}$ & Ref \cite{gusev96:_fluct} & Eqn. (\ref{thin:eqn:stress-stress}) & $Z^{-1} \int C^b_{ij}(z) dz$ \\
\hline
$C_{11}$ & 43.35 & 43.22 & 43.37 \\
$C_{21}$ & 19.01 & 19.45 & 19.22 \\
$C_{44}$ & 22.50 & 22.60 & 22.35 \\
\end{tabular}
\end{ruledtabular}
\caption{Values of the three independent elastic constants of the
bulk fcc crystal in dimensionless Lennard-Jones units. The last
column is the average value of eight atomic layers of the (001)
face in the bulk from Eqn.
(\ref{thin:eqn:local_C}).}\label{thin:data}
\end{table}

The average bulk elastic constants can be calculated from the
local constants using
\begin{equation}\label{thin:eqn:integral}
C^b_{ij} = \frac{1}{Z} \int_Z C^b_{ij}(z) dz,
\end{equation}
where $Z$ is the width of the system. In this example, we set $Z$
to the width of eight atomic layers in the bulk. The average bulk
elastic constants calculated from Eqn. (\ref{thin:eqn:integral})
are given in Table~\ref{thin:data}. The bulk elastic constants
from Eqn. (\ref{thin:eqn:stress-stress}) and the literature
values~\cite{gusev96:_fluct} are also given in
Table~\ref{thin:data}. All three values for each elastic constant
agree well with one another.

\begin{figure}
  \centering
  \includegraphics*[width=\figwidth]{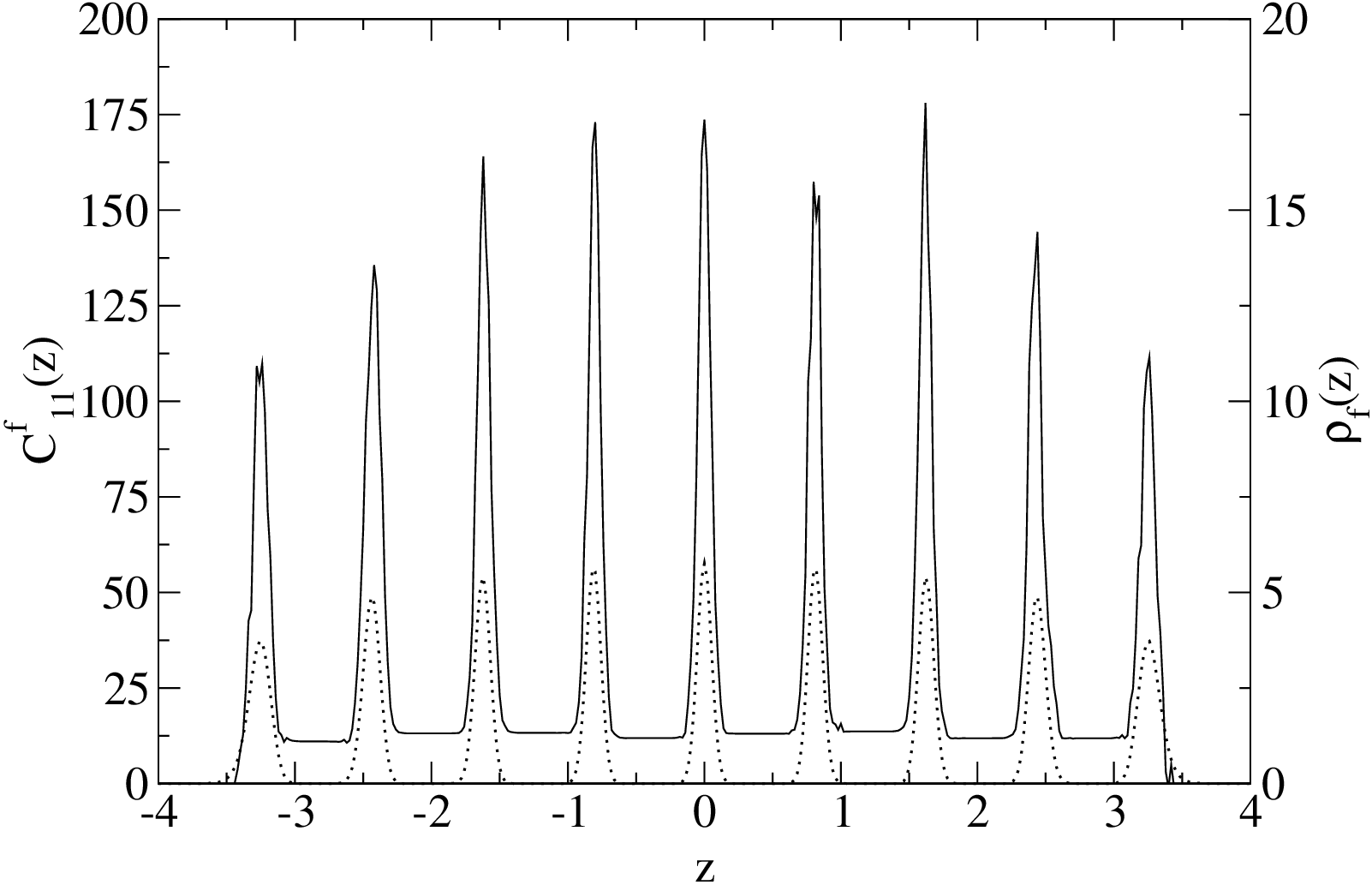}
  \caption{$C_{11}^f(z)$ (solid line) as a function of $z$ for
    the film system from Eqn. (\ref{thin:eqn:local_C}). The density profile,
    $\rho_f(z)$, is shown as the dotted line.}
  \label{thin:fig:C11film}
\end{figure}

\begin{figure}
  \centering
  \includegraphics*[width=\figwidth]{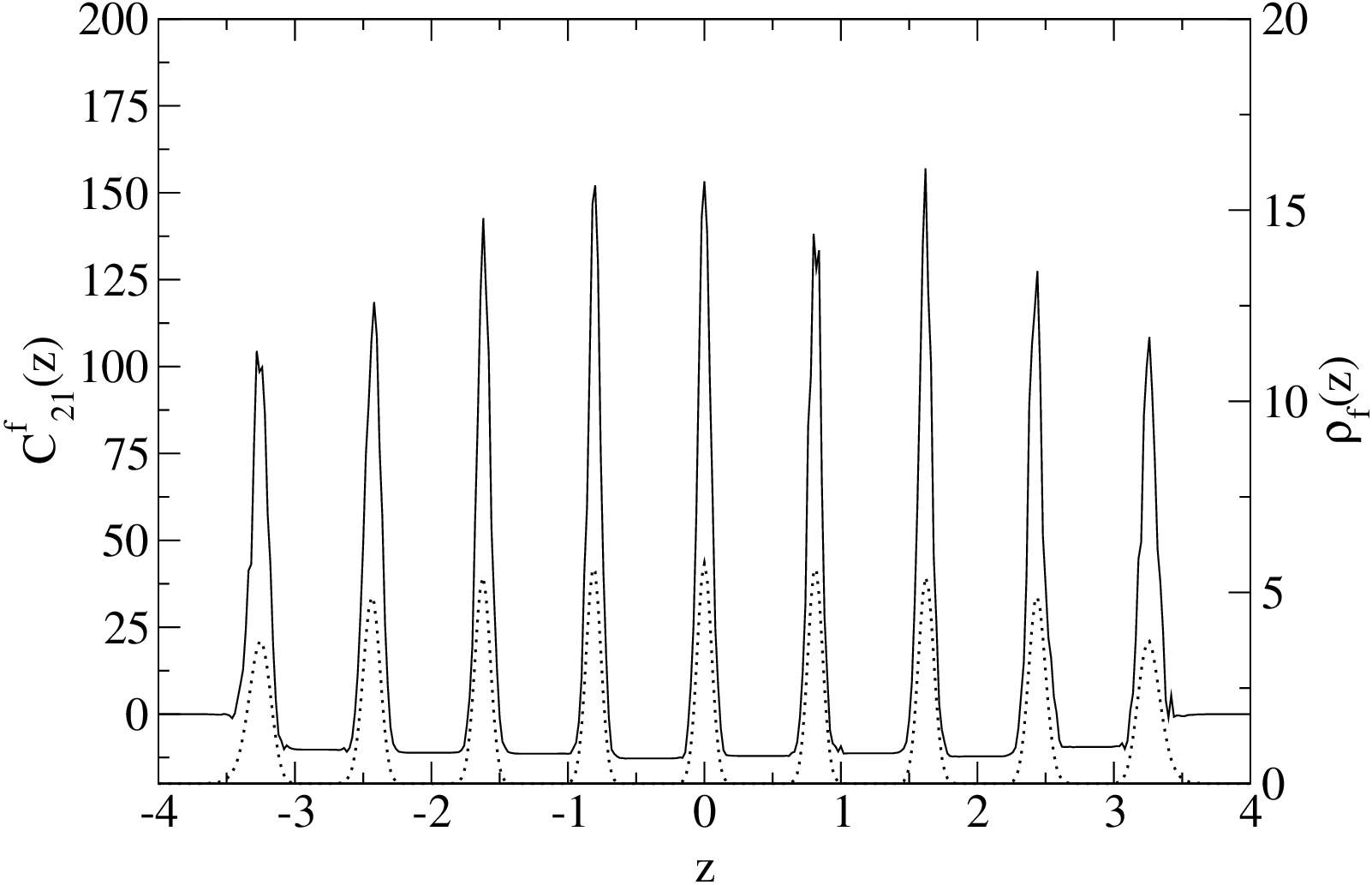}
  \caption{$C_{21}^f(z)$ (solid line) as a function of $z$ for
    the thin film from Eqn. (\ref{thin:eqn:local_C}). The density profile,
    $\rho_f(z)$, is shown as the dotted line.}
  \label{thin:fig:C21film}
\end{figure}

\begin{figure}
  \centering
  \includegraphics*[width=\figwidth]{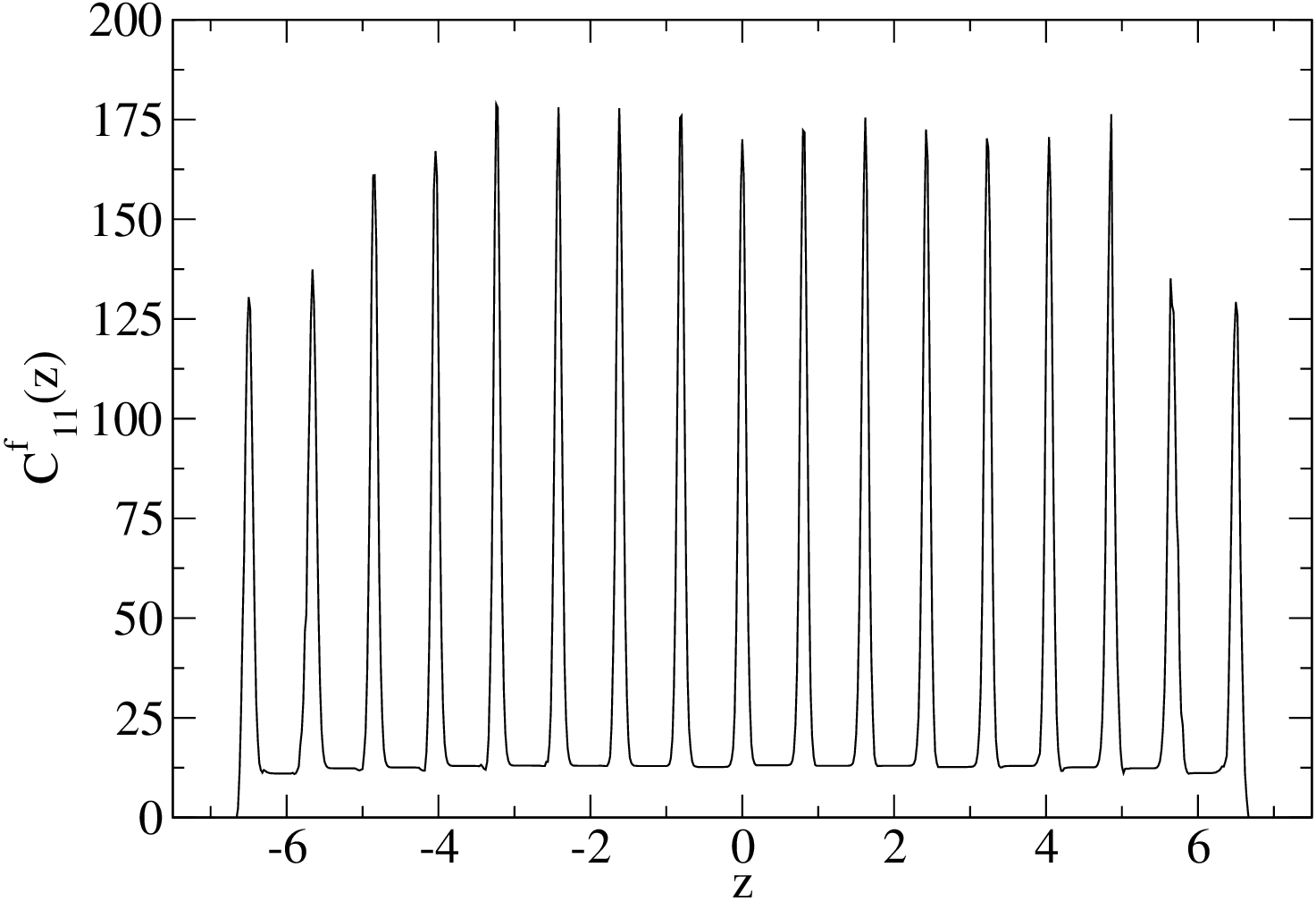}
  \caption{$C_{11}^f(z)$ as a function of $z$ for
    a film with $17$ atomic layers.}
  \label{thin:fig:C11bigfilm}
\end{figure}

The local elastic constant profiles for $C^f_{11}(z)$ and
$C^f_{21}(z)$ in the thin film are shown in
Fig.~\ref{thin:fig:C11film} and Fig.~\ref{thin:fig:C21film},
respectively. The density profile, $\rho_f(z)$, is also shown in
these figures. The peaks corresponding to the atomic layers in the
center of the film ($z=0$) have approximately the same maximum
values as in the bulk system, i.e. $C^f_{ij}(0) \approx
C^b_{ij}(0)$. However, the minimum values between each layer near
the center of the film are less than in the bulk. Interestingly,
$C^f_{21}(z)$ exhibits negative values between each layer. The
meaning of these negative values is discussed below.

The peak values of the elastic constants decrease from the center
of the film as the free surfaces are approached. The profiles also
become broader near the free surfaces. The decrease of the local
elastic constants is an indication of the enhanced atomic mobility
at the surfaces.

In order to investigate the effect of film thickness, a film
consisting of 17 atomic layers was also simulated.
Figure~\ref{thin:fig:C11bigfilm} shows the local elastic constant
profile for $C^f_{11}(z)$ in the film with $17$ layers. The effect
of the free surface is limited to the first two atomic layers for
both this system and that shown in Fig.~\ref{thin:fig:C11film}.
The elastic constant profiles for both film thicknesses are
consistent with one another.

\begin{figure}
  \centering
  \includegraphics*[width=\figwidth]{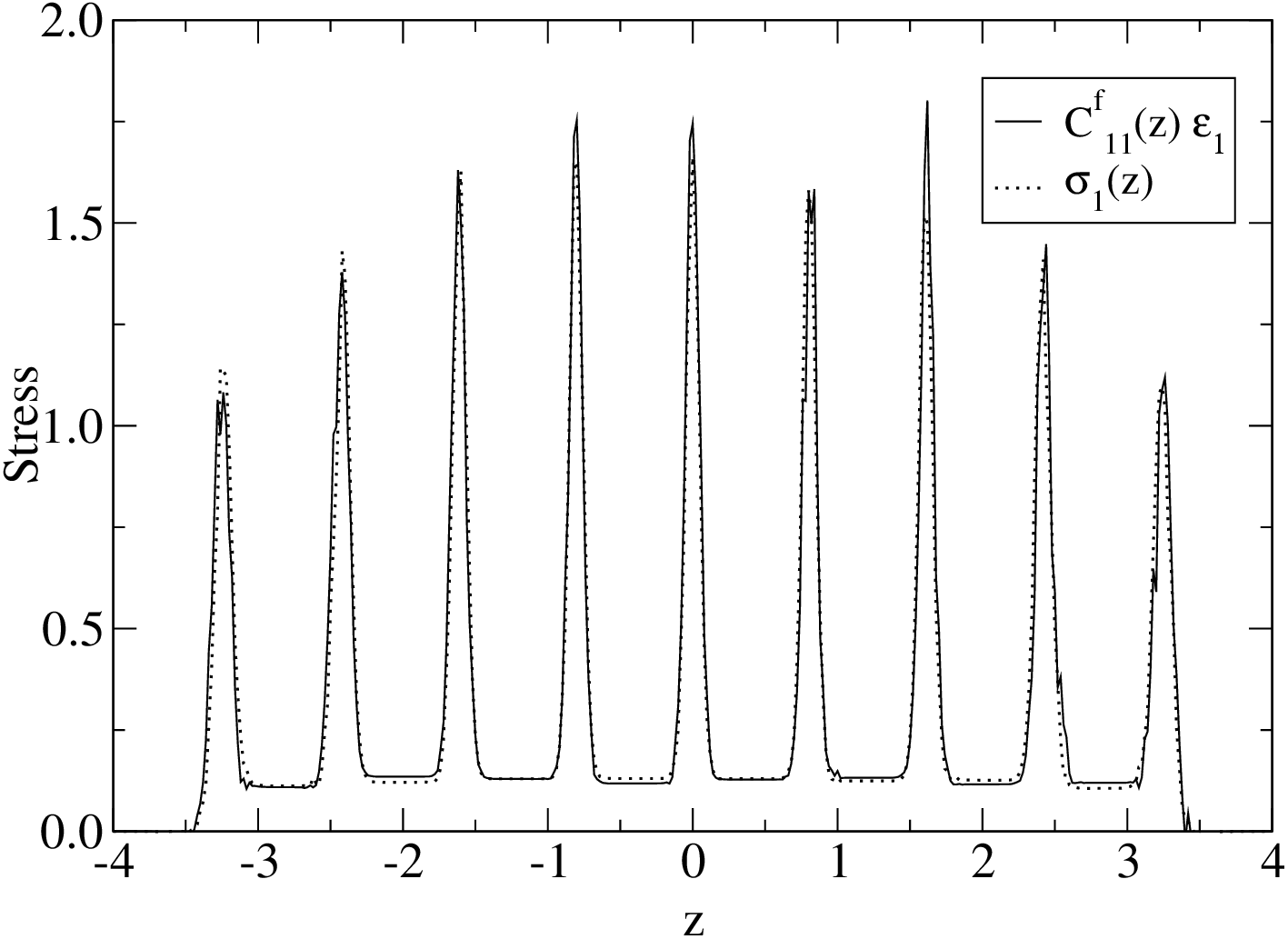}
  \caption{Profiles for $\sigma_1$ in the thin film after a small homogenous uniaxial
  strain, $\epsilon_1$, is applied. The solid line is calculated from the elastic
  constants and Eqn. (\ref{thin:eqn:local_stress}), and the dotted line is calculated directly
  from the simulation using Eqn. (\ref{thin:eqn:MOP_pressure}).}
  \label{thin:fig:s1film}
\end{figure}

\begin{figure}
  \centering
  \includegraphics*[width=\figwidth]{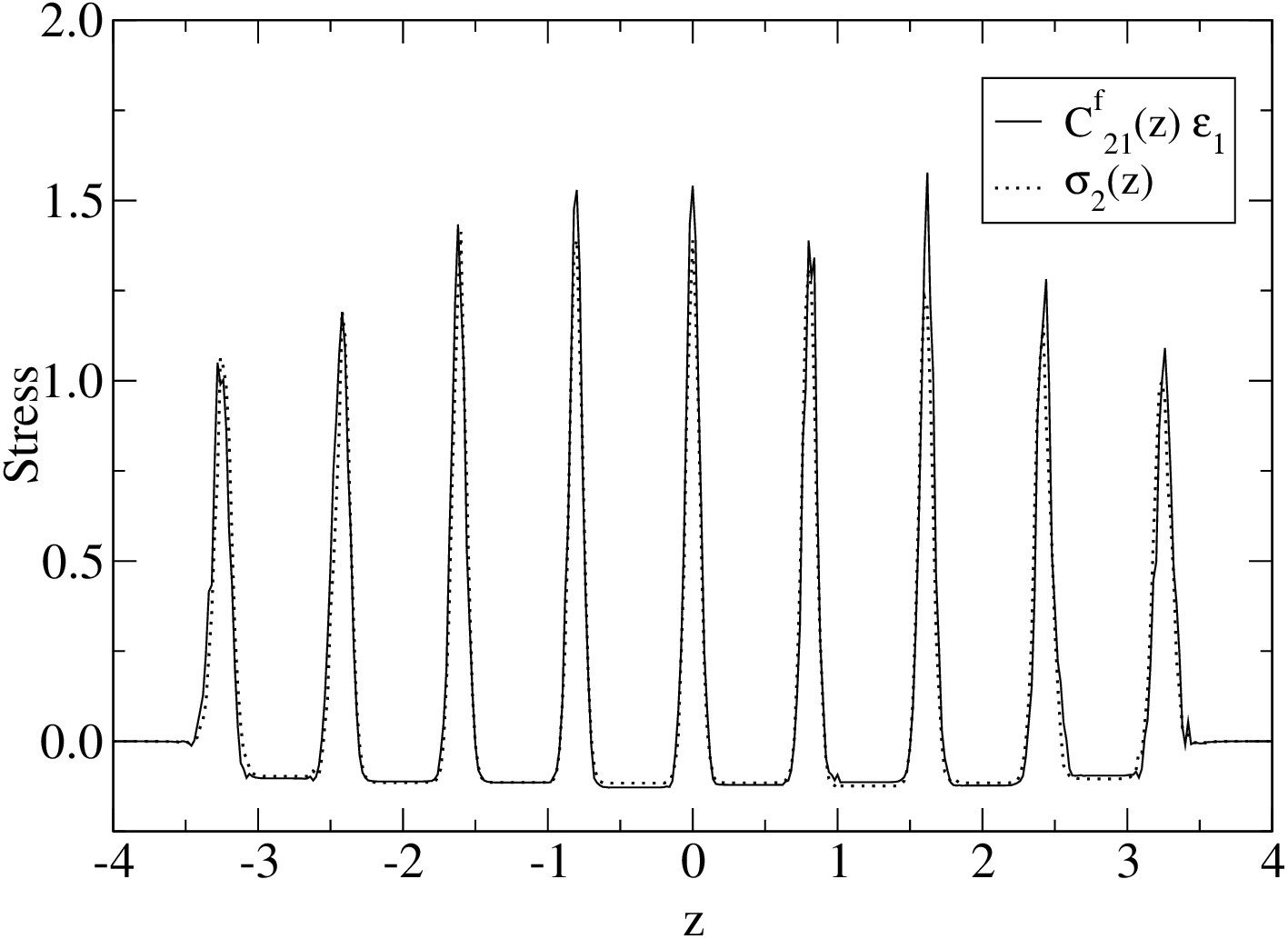}
  \caption{Profiles for $\sigma_2$ in the thin film after a small homogenous uniaxial
  strain, $\epsilon_1$, is applied. The solid line is calculated from the elastic
  constants and Eqn. (\ref{thin:eqn:local_stress}), and the dotted line is calculated directly
  from the simulation using Eqn. (\ref{thin:eqn:MOP_pressure}).}
  \label{thin:fig:s2film}
\end{figure}

The local stress profiles are shown in Fig.~\ref{thin:fig:s1film}
for $\sigma_1(z)$ and in Fig.~\ref{thin:fig:s2film} for
$\sigma_2(z)$ after a homogeneous uniaxial strain was applied. The
local stress profiles in the film were calculated from Eqn.
(\ref{thin:eqn:local_stress}) using the local elasticity tensor
measured at zero strain. The local stress profiles were also
calculated in the strained film as
\begin{equation}\label{thin:eqn:MOP_stress}
\sigma_{i}(z) = - \left[ P_{i}(z)|_{\epsilon=\epsilon_{1}} -
P_{i}(z)|_{\epsilon=0} \right],
\end{equation}
where we used Eqn. (\ref{thin:eqn:MOP_pressure}) for $P_i(z)$. The
results are shown in Fig.~\ref{thin:fig:s1film} and
Fig.~\ref{thin:fig:s2film}. The fact that the two methods for
calculating the local stress profiles give the same result is
reassuring and demonstrates that the response to the applied
strain is linear. For $\sigma_2(z)$, we find that the tensile
(positive) uniaxial strain in $x$-direction causes a negative
stress in the region between the atomic layers
(Fig.~\ref{thin:fig:s2film}). This is directly related to the
negative elastic constants seen in $C^f_{21}(z)$.

\section{Surface Melting}

In order to study the melting behavior of a thin crystalline film,
we adopt the model for argon used by Eerden et
al.~\cite{eerden92:_monte}. As before, we study the (001) surface
of the crystal. The interaction is described by the truncated LJ
potential given by
\begin{equation}
  \begin{split}
  u_{ab} = &4.569\epsilon \left[ \left( \frac{r_{ab}}{\sigma}
  \right)^{-12} - \left( \frac{r_{ab}}{\sigma} \right)^{-6} \right]\\
  &\times \exp \left(
  \frac{0.25\sigma}{r_{ab} - 2.5\sigma} \right).
  \end{split}
\end{equation}
Eerden et al. report the bulk elastic properties for this system.

Consistent with Eerden et al., we have 32 atoms in each layer of
the crystal and use films consisting of 16 layers. The elastic
constants are calculated at planes in the top half of the film
separated by a distance of $dz = 0.02$. At each temperature, the
lateral dimensions of the simulation cell were taken from the
average size of a bulk simulation cell at zero pressure. The
surface of the film was aligned perpendicular to the $z$-axis and
the center of mass was fixed at $z=0$.

Analogous to the definition of the average lateral shear modulus
for a slab between $z$ and $z'$ ($\mu^{\sigma}[z,z']$) by Eerden
et al., we define the local lateral shear modulus at a plane $z$
as
\begin{equation}\label{thin:eqn:amorphous_mu}
  \begin{split}
  \mu^{\sigma}_{\text{i}}(z) = &\frac{1}{8} \sum_{\alpha=x,y} \sum_{\beta=x,y} [
  C_{\alpha \beta \beta \alpha }(z)\\
  &+ C_{\alpha \beta \alpha \beta }(z) - C_{\alpha \alpha \beta \beta }(z) ].
  \end{split}
\end{equation}
Note that this definition is a projected (onto the $xy$-plane)
version of the usual shear modulus for an isotropic
solid~\cite{eerden90}. Since we are interested in the melting
behavior of an anisotropic solid, another useful definition of the
local lateral shear modulus is
\begin{equation}
  \mu^{\sigma}_{\text{a}}(z) = C_{66}(z).
\end{equation}
As the crystal nears its melting point it becomes less anisotropic
and we expect $\mu^{\sigma}_{\text{a}}$ to approach
$\mu^{\sigma}_{\text{i}}$. The melting point is defined here as
the temperature at which $\mu^{\sigma}_{\text{a}}$ and
$\mu^{\sigma}_{\text{i}}$ vanish.

\begin{figure}
  \centering
  \includegraphics*[width=\figwidth]{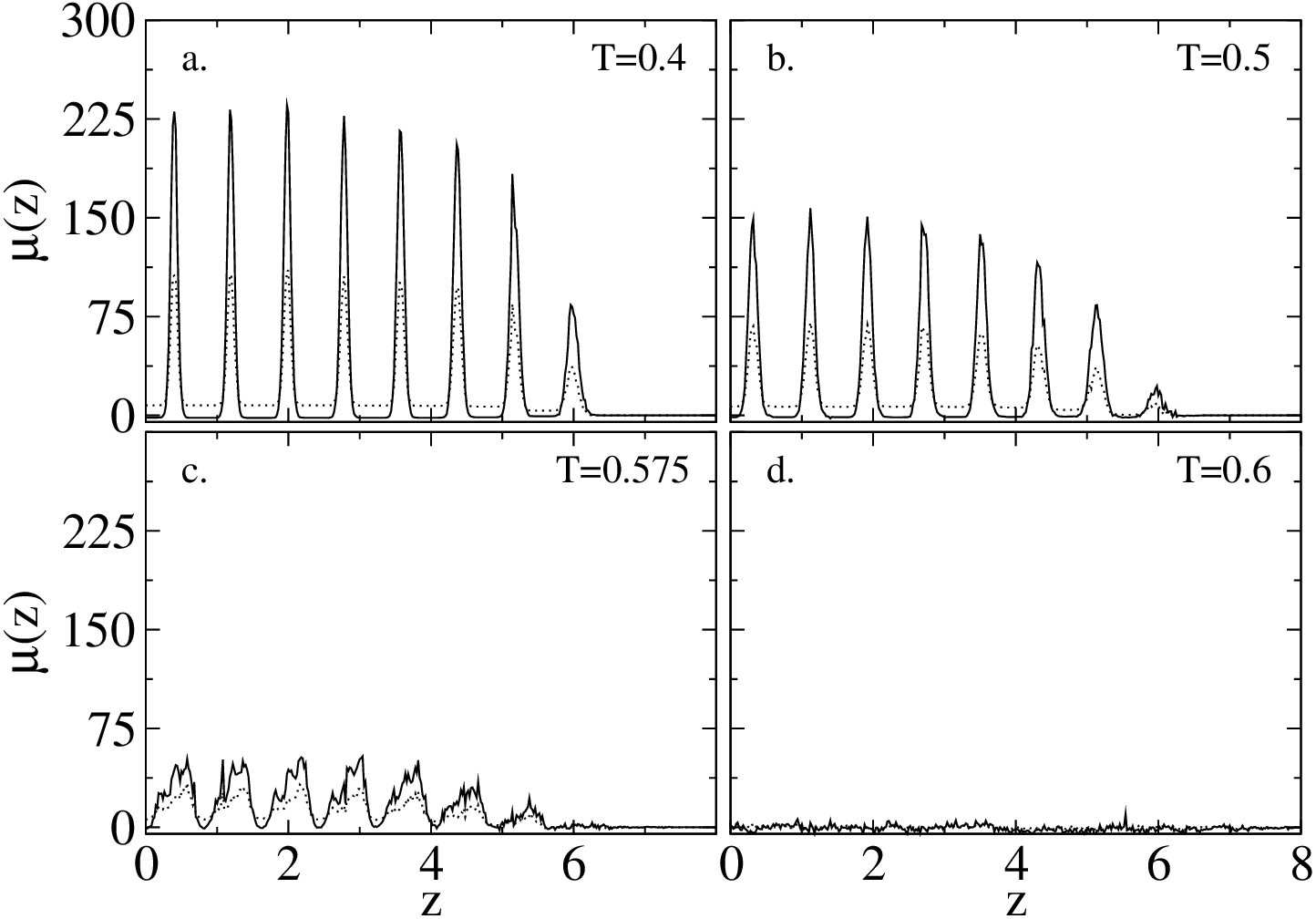}
  \caption{The shear moduli profiles in a thin film of argon at four
  temperatures. The solid line is the $\mu^{\sigma}_{\text{a}}$ and the dotted line
  is $\mu^{\sigma}_{\text{i}}$.}
  \label{thin:fig:eerden_profile}
\end{figure}

\begin{figure}
  \centering
  \includegraphics*[width=\figwidth]{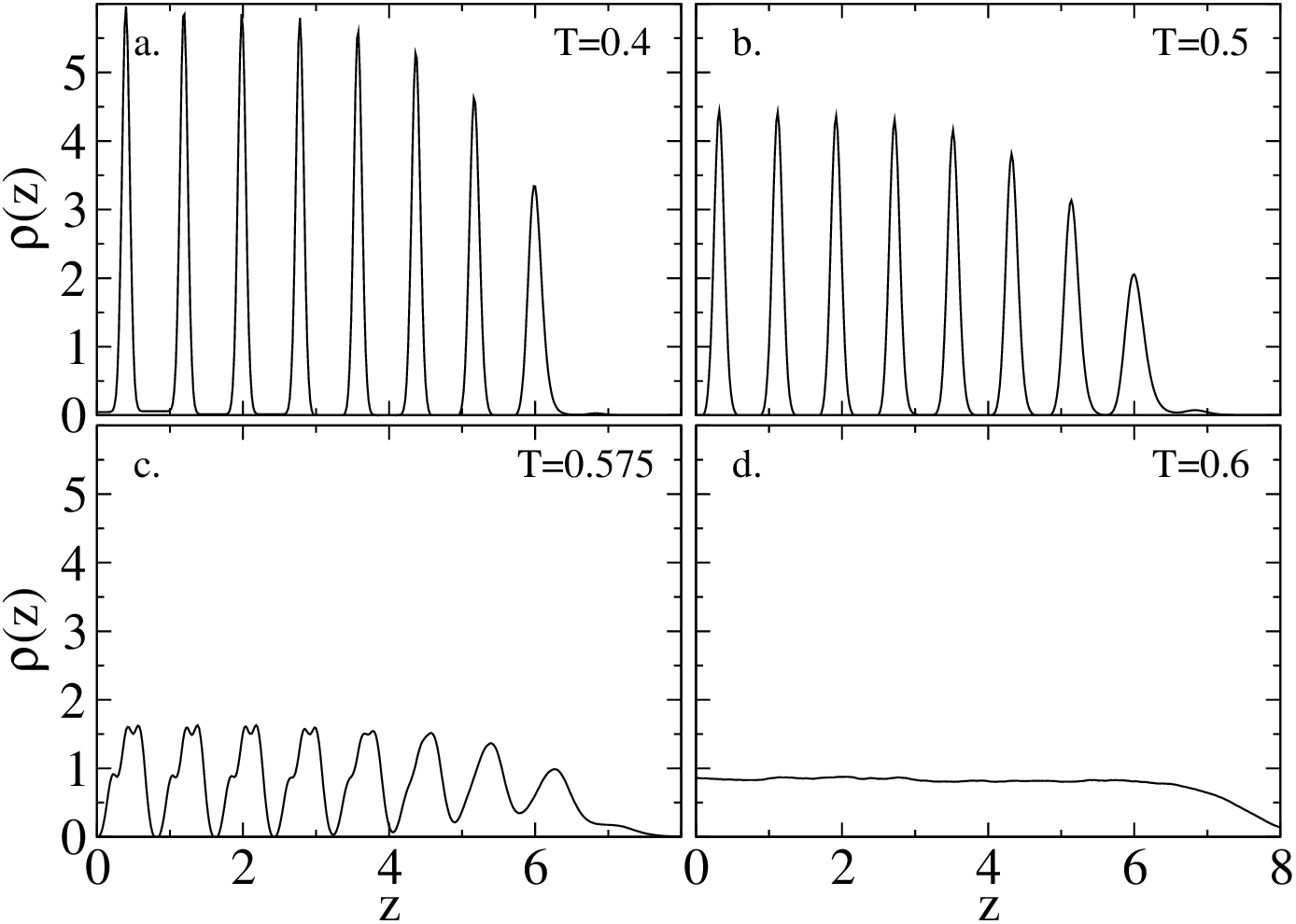}
  \caption{The density profiles in a thin film of argon at four
  temperatures.}
  \label{thin:fig:density}
\end{figure}

The shear moduli as a function of position in the film are shown
in Fig.~\ref{thin:fig:eerden_profile} at four different
temperatures. The density profiles at these temperatures are shown
in Fig.~\ref{thin:fig:density}. The density profile has units of
$\sigma^{-3}$ and its integral over the entire system, $A \int_z
\rho(z) dz$, gives the total number of particles in the film. In
the following discussion, we will refer to the layers starting
with the surface layer as layer-1, layer-2, etc.

The behavior at the lowest temperature
(Fig.~\ref{thin:fig:eerden_profile}a.), $T=0.4$, is similar to
that of the NNLJ film, having bulk behavior in the center of the
film and decreasing moduli in the layers near the surface. The
difference between $\mu^{\sigma}_{\text{i}}$ and
$\mu^{\sigma}_{\text{a}}$ reflects the fact that the crystal is
anisotropic, even in the layer nearest to the surface, layer-1.
This is also evident in the density profile
(Fig.~\ref{thin:fig:density}a.) where all the atomic layers of the
crystal are separated by regions of empty space. At temperatures
above $T=0.4$, isolated atoms have sufficient energy to escape
layer-1 and occupy positions outside the film (layer-0). This
additional layer can be seen as the very small peak centered at
$z=6.82$. The additional layer, however, has a zero shear modulus.

At $T=0.5$, the shear moduli
(Fig.~\ref{thin:fig:eerden_profile}b.) of each layer have
decreased from those at $T=0.4$, indicating a softening of the
crystal. The difference between $\mu^{\sigma}_{\text{i}}$ and
$\mu^{\sigma}_{\text{a}}$ has decreased considerably in layer-1,
indicating nearly isotropic behavior near the melting temperature.
The number of atoms which escape from layer-1 has increased,
indicated by the larger peak or shoulder in the density profile
(Fig.~\ref{thin:fig:density}b.) at $z=6.82$.

In Fig.~\ref{thin:fig:eerden_profile}c. and
Fig.~\ref{thin:fig:density}c. ($T=0.575$), the behavior of the
layers near the surface has changed significantly. Both
$\mu^{\sigma}_{\text{i}}$ and $\mu^{\sigma}_{\text{a}}$ are
essentially zero (indicating isotropy and melting) at layer-1 even
though the density profile exhibits some remaining structure in
that region. Between layer-1 and layer-2 and between layer-2 and
layer-3, the density is non-zero yet the shear modulus is zero. A
small amount of argon exists as a fluid between these layers.
Layer-0 contains even more atoms at this temperature and shows a
flat density shoulder which decays to zero, indicating a loss of
structure at the film-vacuum interface. At temperatures just below
$T=0.575$ and above, the delineation of layer thickness becomes
ambiguous and the use of a layer-averaged shear modulus becomes
questionable. The method of planes proposed here eliminates that
ambiguity.

At $T=0.6$, the shear modulus of the entire film is zero and the
density profile is flat. The film is a liquid throughout and has
none of the structure originally present in the crystalline film
at lower temperatures.

\begin{figure}
  \centering
  \includegraphics*[width=\figwidth]{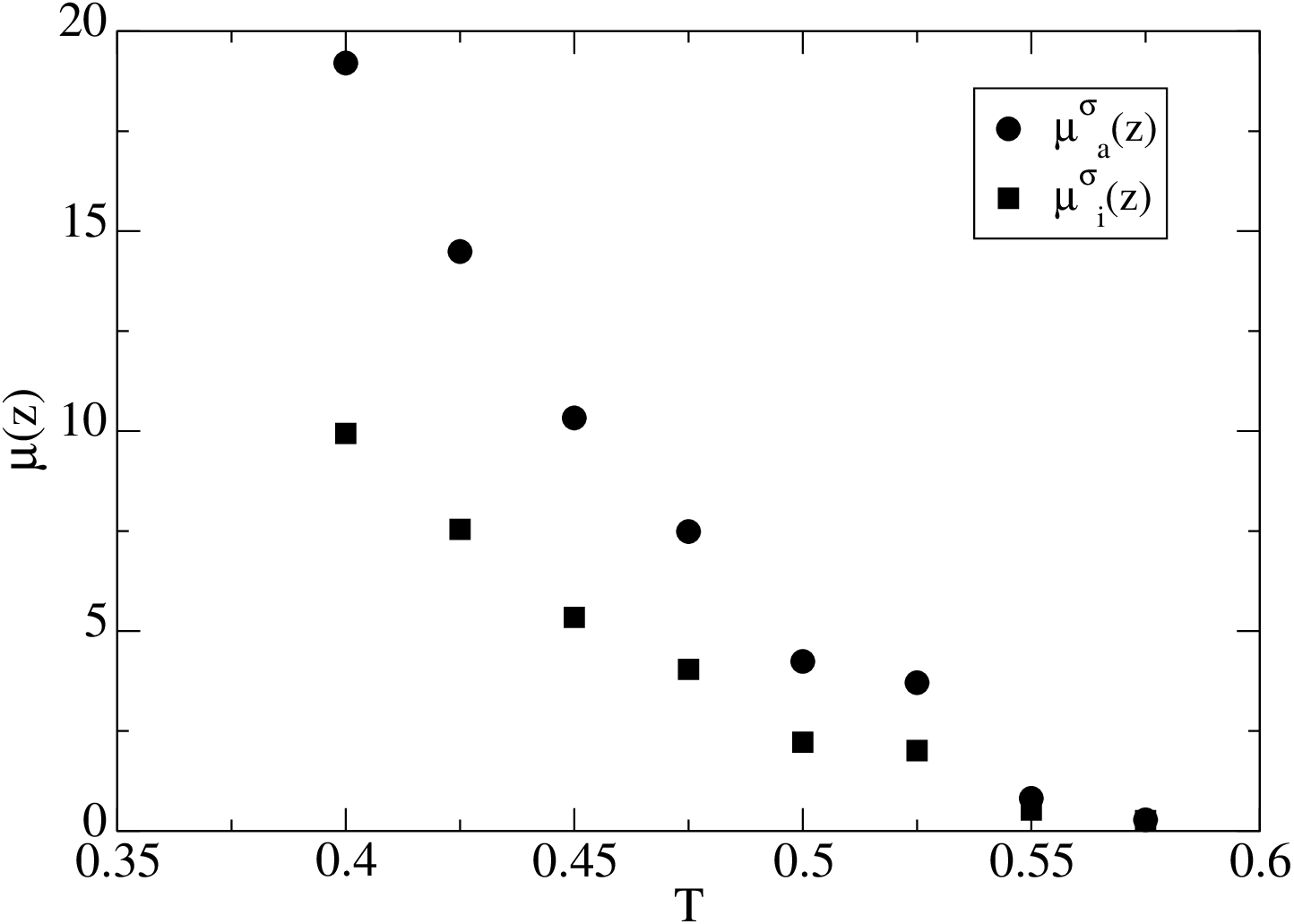}
  \caption{The average lateral shear moduli in a thin argon film
  as a function of temperature.}
  \label{thin:fig:surfacemu}
\end{figure}

An average shear modulus for the surface layer can be defined by
integrating the profiles in Fig.~\ref{thin:fig:eerden_profile}.
The average shear modulus is given by
\begin{equation}\label{thin:eqn:avgsurface}
  \overline{\mu}^{\sigma} = \frac{1}{\Delta z}
  \int_{z_{\text{min}}}^{z_{\text{max}}}
  \mu^{\sigma}(z) dz.
\end{equation}
A layer thickness $\Delta z$ must first be defined in order to
perform the integration. We arbitrarily choose $\Delta z$ for the
surface layer to be the distance between the peaks in the density
profile of layer-1 and layer-2 at each temperature. In Eqn.
({\ref{thin:eqn:avgsurface}), $z_{\text{min}}$ is the location of
the minimum density between layer-1 and layer-2 and
$z_{\text{max}} = z_{\text{min}} + \Delta z$.

The results for $\overline{\mu}^{\sigma}_{\text{i}}$ and
$\overline{\mu}^{\sigma}_{\text{a}}$ of the surface layer are
shown in Fig.~\ref{thin:fig:surfacemu} for temperatures up to
$T=0.575$. Both shear moduli decrease sharply with increasing
temperature and vanish at $T=0.575$, indicating melting of the
surface layer. This is in agreement with the literature value of
$T=0.576$. The bulk melting temperature is
$T_b=0.601$~\cite{eerden92:_monte}.

\section{Conclusion}

A formalism for calculation of the local elastic constants in
inhomogeneous systems based on the method of planes has been
presented. Unlike previous work, this formalism does not require
the partitioning of the system into a set of finite volumes or
``slabs'' over which average elastic constants are calculated. As
an example application of the technique, Monte Carlo simulations
of the nearest-neighbor Lennard-Jones fcc crystal in the bulk and
in thin film geometries have been presented.

The local atomic structure of the crystals was evident in the
local elastic constants calculated at precise planes. In the thin
film, the elastic constants are decreased from the corresponding
bulk values, especially near the free surfaces. This decrease near
a free surface is expected to give rise to apparent deviations
from bulk continuum behavior in thin films and nanoscopic
structures.

The melting behavior of argon in a thin film was also investigated
within the context of this formalism. Results show how the shear
modulus profile of the surface layer of atoms vanishes below the
melting temperature of the core of the film. Below the melting
temperature of the film, the free surface allows sufficient
thermal motion for the surface atoms to reach an isotropic liquid
state prior to the bulk of the film.


\begin{thebibliography}{26}
\expandafter\ifx\csname
natexlab\endcsname\relax\def\natexlab#1{#1}\fi
\expandafter\ifx\csname bibnamefont\endcsname\relax
  \def\bibnamefont#1{#1}\fi
\expandafter\ifx\csname bibfnamefont\endcsname\relax
  \def\bibfnamefont#1{#1}\fi
\expandafter\ifx\csname citenamefont\endcsname\relax
  \def\citenamefont#1{#1}\fi
\expandafter\ifx\csname url\endcsname\relax
  \def\url#1{\texttt{#1}}\fi
\expandafter\ifx\csname
urlprefix\endcsname\relax\def\urlprefix{URL }\fi
\providecommand{\bibinfo}[2]{#2}
\providecommand{\eprint}[2][]{\url{#2}}

\bibitem[{\citenamefont{Tanaka et~al.}(1993)\citenamefont{Tanaka, Morigami, and
  Atoda}}]{tanaka93}
\bibinfo{author}{\bibfnamefont{T.}~\bibnamefont{Tanaka}},
  \bibinfo{author}{\bibfnamefont{M.}~\bibnamefont{Morigami}}, \bibnamefont{and}
  \bibinfo{author}{\bibfnamefont{N.}~\bibnamefont{Atoda}},
  \bibinfo{journal}{Jpn. J. Appl. Phys.} \textbf{\bibinfo{volume}{32}},
  \bibinfo{pages}{6059} (\bibinfo{year}{1993}).

\bibitem[{\citenamefont{Cao et~al.}(2000)\citenamefont{Cao, Nealey, and
  Domke}}]{cao00}
\bibinfo{author}{\bibfnamefont{H.}~\bibnamefont{Cao}},
  \bibinfo{author}{\bibfnamefont{P.}~\bibnamefont{Nealey}}, \bibnamefont{and}
  \bibinfo{author}{\bibfnamefont{W.}~\bibnamefont{Domke}}, \bibinfo{journal}{J.
  Vac. Sci. \& Tech. B} \textbf{\bibinfo{volume}{18}}, \bibinfo{pages}{3303}
  (\bibinfo{year}{2000}).

\bibitem[{\citenamefont{Boehme and de~Pablo}(2002)}]{boehme02}
\bibinfo{author}{\bibfnamefont{T.}~\bibnamefont{Boehme}} \bibnamefont{and}
  \bibinfo{author}{\bibfnamefont{J.}~\bibnamefont{de~Pablo}},
  \bibinfo{journal}{J. Chem. Phys.} \textbf{\bibinfo{volume}{116}},
  \bibinfo{pages}{9939} (\bibinfo{year}{2002}).

\bibitem[{\citenamefont{Fryer et~al.}(2001)\citenamefont{Fryer, Peters, Kim,
  Tomaszewski, {de Pablo}, Nealey, White, and Wu}}]{fryer2}
\bibinfo{author}{\bibfnamefont{D.}~\bibnamefont{Fryer}},
  \bibinfo{author}{\bibfnamefont{R.}~\bibnamefont{Peters}},
  \bibinfo{author}{\bibfnamefont{E.}~\bibnamefont{Kim}},
  \bibinfo{author}{\bibfnamefont{J.}~\bibnamefont{Tomaszewski}},
  \bibinfo{author}{\bibfnamefont{J.~J.} \bibnamefont{{de Pablo}}},
  \bibinfo{author}{\bibfnamefont{P.}~\bibnamefont{Nealey}},
  \bibinfo{author}{\bibfnamefont{C.}~\bibnamefont{White}}, \bibnamefont{and}
  \bibinfo{author}{\bibfnamefont{W.}~\bibnamefont{Wu}},
  \bibinfo{journal}{Macromolecules} \textbf{\bibinfo{volume}{34}},
  \bibinfo{pages}{5627} (\bibinfo{year}{2001}).

\bibitem[{\citenamefont{Torres et~al.}(2000)\citenamefont{Torres, Nealey, and
  {de Pablo}}}]{torres00}
\bibinfo{author}{\bibfnamefont{J.~A.} \bibnamefont{Torres}},
  \bibinfo{author}{\bibfnamefont{P.~F.} \bibnamefont{Nealey}}, \bibnamefont{and}
  \bibinfo{author}{\bibfnamefont{J.~J.} \bibnamefont{{de Pablo}}},
  \bibinfo{journal}{Phys. Rev. Lett.} \textbf{\bibinfo{volume}{85}},
  \bibinfo{pages}{3221} (\bibinfo{year}{2000}).

\bibitem[{\citenamefont{Lutsko}(1988)}]{lutsko88}
\bibinfo{author}{\bibfnamefont{J.}~\bibnamefont{Lutsko}}, \bibinfo{journal}{J.
  Appl. Phys.} \textbf{\bibinfo{volume}{64}}, \bibinfo{pages}{1152}
  (\bibinfo{year}{1988}).

\bibitem[{\citenamefont{van~der Eerden et~al.}(1990)\citenamefont{van~der
  Eerden, Roos, and van~der Veer}}]{eerden90}
\bibinfo{author}{\bibfnamefont{J.}~\bibnamefont{van~der Eerden}},
  \bibinfo{author}{\bibfnamefont{A.}~\bibnamefont{Roos}}, \bibnamefont{and}
  \bibinfo{author}{\bibfnamefont{J.}~\bibnamefont{van~der Veer}},
  \bibinfo{journal}{J. Crystal Growth} \textbf{\bibinfo{volume}{99}},
  \bibinfo{pages}{77} (\bibinfo{year}{1990}).

\bibitem[{\citenamefont{Ray}(1982)}]{ray82:_fluct}
\bibinfo{author}{\bibfnamefont{J.~R.} \bibnamefont{Ray}}, \bibinfo{journal}{J.
  Appl. Phys.} \textbf{\bibinfo{volume}{53}}, \bibinfo{pages}{6441}
  (\bibinfo{year}{1982}).

\bibitem[{\citenamefont{Todd et~al.}(1995)\citenamefont{Todd, Evans, and
  Daivis}}]{todd95:_press}
\bibinfo{author}{\bibfnamefont{B.~D.} \bibnamefont{Todd}},
  \bibinfo{author}{\bibfnamefont{D.~J.} \bibnamefont{Evans}}, \bibnamefont{and}
  \bibinfo{author}{\bibfnamefont{P.~J.} \bibnamefont{Daivis}},
  \bibinfo{journal}{Phys. Rev. E} \textbf{\bibinfo{volume}{52}},
  \bibinfo{pages}{1627} (\bibinfo{year}{1995}).

\bibitem[{\citenamefont{Varnik et~al.}(2000)\citenamefont{Varnik, Baschnagel,
  and Binder}}]{varnik00:_molec}
\bibinfo{author}{\bibfnamefont{F.}~\bibnamefont{Varnik}},
  \bibinfo{author}{\bibfnamefont{J.}~\bibnamefont{Baschnagel}},
  \bibnamefont{and} \bibinfo{author}{\bibfnamefont{K.}~\bibnamefont{Binder}},
  \bibinfo{journal}{J. Chem. Phys.} \textbf{\bibinfo{volume}{113}},
  \bibinfo{pages}{4444} (\bibinfo{year}{2000}).

\bibitem[{\citenamefont{Irving and Kirkwood}(1950)}]{irving50}
\bibinfo{author}{\bibfnamefont{J.~H.} \bibnamefont{Irving}} \bibnamefont{and}
  \bibinfo{author}{\bibfnamefont{J.~G.} \bibnamefont{Kirkwood}},
  \bibinfo{journal}{J. Chem. Phys.} \textbf{\bibinfo{volume}{18}},
  \bibinfo{pages}{817} (\bibinfo{year}{1950}).

\bibitem[{\citenamefont{Tsai}(1979)}]{tsai79}
\bibinfo{author}{\bibfnamefont{D.}~\bibnamefont{Tsai}}, \bibinfo{journal}{J.
  Chem. Phys.} \textbf{\bibinfo{volume}{70}}, \bibinfo{pages}{1275}
  (\bibinfo{year}{1979}).

\bibitem[{\citenamefont{Rowlinson and Widom}(1982)}]{rowlinson82}
\bibinfo{author}{\bibfnamefont{J.}~\bibnamefont{Rowlinson}} \bibnamefont{and}
  \bibinfo{author}{\bibfnamefont{B.}~\bibnamefont{Widom}},
  \emph{\bibinfo{title}{Molecular Therory of Capillarity}}
  (\bibinfo{publisher}{Oxford}, \bibinfo{year}{1982}).

\bibitem[{\citenamefont{Harasima}(1958)}]{harasima58}
\bibinfo{author}{\bibfnamefont{A.}~\bibnamefont{Harasima}},
  \bibinfo{journal}{Adv. Chem. Phys.}  (\bibinfo{year}{1958}).

\bibitem[{\citenamefont{Hafskjold and Ikeshoji}(2002)}]{hafskjold02}
\bibinfo{author}{\bibfnamefont{B.}~\bibnamefont{Hafskjold}} \bibnamefont{and}
  \bibinfo{author}{\bibfnamefont{T.}~\bibnamefont{Ikeshoji}},
  \bibinfo{journal}{Phys. Rev. E} \textbf{\bibinfo{volume}{66}},
  \bibinfo{pages}{011203} (\bibinfo{year}{2002}).

\bibitem[{\citenamefont{Parrinello and Rahman}(1981)}]{parrinello81:_polym}
\bibinfo{author}{\bibfnamefont{M.}~\bibnamefont{Parrinello}} \bibnamefont{and}
  \bibinfo{author}{\bibfnamefont{A.}~\bibnamefont{Rahman}},
  \bibinfo{journal}{J. Appl. Phys.} \textbf{\bibinfo{volume}{52}},
  \bibinfo{pages}{7182} (\bibinfo{year}{1981}).

\bibitem[{\citenamefont{Fay and Ray}(1992)}]{fay92:_monte}
\bibinfo{author}{\bibfnamefont{P.~J.} \bibnamefont{Fay}} \bibnamefont{and}
  \bibinfo{author}{\bibfnamefont{J.~R.} \bibnamefont{Ray}},
  \bibinfo{journal}{Phys. Rev. A} \textbf{\bibinfo{volume}{46}},
  \bibinfo{pages}{4645} (\bibinfo{year}{1992}).

\bibitem[{\citenamefont{Sprik et~al.}(1984)\citenamefont{Sprik, Impey, and
  Klein}}]{sprik84:_secon}
\bibinfo{author}{\bibfnamefont{M.}~\bibnamefont{Sprik}},
  \bibinfo{author}{\bibfnamefont{R.~W.} \bibnamefont{Impey}}, \bibnamefont{and}
  \bibinfo{author}{\bibfnamefont{M.~L.} \bibnamefont{Klein}},
  \bibinfo{journal}{Phys. Rev. B} \textbf{\bibinfo{volume}{29}},
  \bibinfo{pages}{4368} (\bibinfo{year}{1984}).

\bibitem[{\citenamefont{Cowley}(1983)}]{cowley83:_some}
\bibinfo{author}{\bibfnamefont{E.~R.} \bibnamefont{Cowley}},
  \bibinfo{journal}{Phys. Rev. B} \textbf{\bibinfo{volume}{28}},
  \bibinfo{pages}{3160} (\bibinfo{year}{1983}).

\bibitem[{\citenamefont{Li and Johnson}(1992)}]{li92:_fluct}
\bibinfo{author}{\bibfnamefont{M.}~\bibnamefont{Li}} \bibnamefont{and}
  \bibinfo{author}{\bibfnamefont{W.~L.} \bibnamefont{Johnson}},
  \bibinfo{journal}{Phys. Rev. B} \textbf{\bibinfo{volume}{46}},
  \bibinfo{pages}{5237} (\bibinfo{year}{1992}).

\bibitem[{\citenamefont{Ray et~al.}(1985)\citenamefont{Ray, Moody, and
  Rahman}}]{ray85:_molec}
\bibinfo{author}{\bibfnamefont{J.~R.} \bibnamefont{Ray}},
  \bibinfo{author}{\bibfnamefont{M.~C.} \bibnamefont{Moody}}, \bibnamefont{and}
  \bibinfo{author}{\bibfnamefont{A.}~\bibnamefont{Rahman}},
  \bibinfo{journal}{Phys. Rev. B} \textbf{\bibinfo{volume}{32}},
  \bibinfo{pages}{733} (\bibinfo{year}{1985}).

\bibitem[{\citenamefont{Nye}(1984)}]{nye84:_physic}
\bibinfo{author}{\bibfnamefont{J.~F.} \bibnamefont{Nye}},
  \emph{\bibinfo{title}{Physical Properties of Crystals: Their Representation
  by Tensors and Matrices}} (\bibinfo{publisher}{Clarendon Press, Oxford
  University Press}, \bibinfo{year}{1984}).

\bibitem[{\citenamefont{Love}(1927)}]{love27}
\bibinfo{author}{\bibfnamefont{A.~E.~H.} \bibnamefont{Love}},
  \emph{\bibinfo{title}{A Treastise on the Mathematical Theory of Elasticity}}
  (\bibinfo{publisher}{Cambridge Univ. Press.}, \bibinfo{year}{1927}),
  \bibinfo{edition}{4th} ed.

\bibitem[{\citenamefont{Kluge et~al.}(1990)\citenamefont{Kluge, Wolf, Lutsko,
  and Phillpot}}]{kluge90}
\bibinfo{author}{\bibfnamefont{M.}~\bibnamefont{Kluge}},
  \bibinfo{author}{\bibfnamefont{D.}~\bibnamefont{Wolf}},
  \bibinfo{author}{\bibfnamefont{J.}~\bibnamefont{Lutsko}}, \bibnamefont{and}
  \bibinfo{author}{\bibfnamefont{S.}~\bibnamefont{Phillpot}},
  \bibinfo{journal}{J. Appl. Phys.} \textbf{\bibinfo{volume}{67}},
  \bibinfo{pages}{2370} (\bibinfo{year}{1990}).

\bibitem[{\citenamefont{Gusev et~al.}(1996)\citenamefont{Gusev, Zehnder, and
  Suter}}]{gusev96:_fluct}
\bibinfo{author}{\bibfnamefont{A.~A.} \bibnamefont{Gusev}},
  \bibinfo{author}{\bibfnamefont{M.~M.} \bibnamefont{Zehnder}},
  \bibnamefont{and} \bibinfo{author}{\bibfnamefont{U.~W.} \bibnamefont{Suter}},
  \bibinfo{journal}{Phys. Rev. B} \textbf{\bibinfo{volume}{54}},
  \bibinfo{pages}{1} (\bibinfo{year}{1996}).

\bibitem[{\citenamefont{v.d. Eerden et~al.}(1992)\citenamefont{v.d. Eerden,
  Knops, and Roos}}]{eerden92:_monte}
\bibinfo{author}{\bibfnamefont{J.~P.} \bibnamefont{v.d. Eerden}},
  \bibinfo{author}{\bibfnamefont{H.~J.~F.} \bibnamefont{Knops}},
  \bibnamefont{and} \bibinfo{author}{\bibfnamefont{A.}~\bibnamefont{Roos}},
  \bibinfo{journal}{J. Chem. Phys.} \textbf{\bibinfo{volume}{96}},
  \bibinfo{pages}{714} (\bibinfo{year}{1992}).

\end{thebibliography}
\end{document}